\documentclass[11pt,a4paper]{article}
\usepackage{latexsym,a4}

\newcommand{\op}[1]{\ensuremath{\widehat{\textsf{\ensuremath{#1}}}}}

\newcommand{\ket}[1]{\left | \, #1 \right \rangle}

\newcommand{\bra}[1]{\left \langle \, #1 \right |}
\newcommand{\be}{\begin{equation}}
\newcommand{\ee}{\end{equation}}
\newcommand{\BE}{\begin{eqnarray}}
\newcommand{\EE}{\end{eqnarray}}
\newcommand{\bq}{\begin{quote}}
\newcommand{\eq}{\end{quote}}
\newcommand{\nn}{\nonumber}
\newcommand{\forget}[1]{}
\newcommand{\up}{\uparrow}
\newcommand{\down}{\downarrow}

\newcommand{\Tr}{\mbox{Tr~}}

\begin{document}
\title{Sufficient conditions for three-particle entanglement and
their  tests in recent experiments}
\author{
Michael Seevinck\\[.2mm]
\emph{\small Sub-Faculty of Physics, University of Nijmegen}\\
\emph{\small PO Box 9010, 6500 GL Nijmegen, the Netherlands,}\\
\emph{\small Institute for History and Foundations of Science,}\\
\emph{\small University of Utrecht, PO Box 80.000, 3508 TA
Utrecht, The Netherlands}\\
{\small \tt michielp@sci.kun.nl}
\\[.2 mm]
and \\[.6mm]
Jos Uffink\\
%\emph{
\emph{ \small Institute for History and Foundations of Science,}\\
\emph{ \small University of Utrecht, PO Box 80.000, 3508 TA
Utrecht, the Netherlands,}\\
  {\small \tt uffink@phys.uu.nl}
}
 \maketitle
\begin{abstract}
We point out a loophole problem in some recent experimental claims
to produce three-particle entanglement. The problem consists in
the question whether mixtures of two-particle entangled states
might suffice to explain the experimental data.
 In an attempt to close this loophole,
 we review two sufficient conditions that distinguish between
  $N$-particle states in which all $N$ particles are entangled to
  each other and states in which only $M$ particles are entangled (with $M<N$).
  It is shown that three recent experiments to obtain three-particle
  entangled states
  (Bouwmeester \emph{et al.}~Phys.\ Rev.\ Lett.\ {\bf 82}, 1345 (1999),
  Pan \emph{et al.}~Nature {\bf 403}, 515 (2000), and Rauschenbeutel
  \emph{et al.} Science {\bf 288}, 2024, (2000))
  do not meet these conditions. We conclude that the question
whether these experiments provide confirmation of three-particle
entanglement remains unresolved.  We also propose modifications of
the experiments that would make such
confirmation feasible.\\[.2cm]
\end{abstract}

\section{Introduction} \normalsize The experimental production
and detection of multiparticle entanglement has seen much
progress during the last years. Manipulation of such highly
entangled $N$-particle states is of great interest for
implementing quantum information techniques, such as quantum
computing and quantum cryptography, as well as for fundamental
tests of quantum mechanics. Extended efforts have resulted in
recent claims of experimental confirmation of both three- and
four-particle entanglement using photons and atom-cavity
techniques [1--5]\nocite{BOU1998,RAUSCH,SACKETT,PAN2}. In this
paper, we examine a possible loophole in such claims.

$N$-particle entanglement differs from the more well-known
two-particle entanglement, not only because the classification of
different types of this form of entanglement is still an open
problem \cite{POPESCU, LEWENSTEIN}, but also because it  requires
different conditions for actual experimental confirmation. In the
case of two-particle entangled states, it suffices to show that
the observed data  cannot be explained by  a ``local realist''
model. That is, it is sufficient for the correlations between the
observed data to violate a certain Bell inequality. In fact, for
pure states, this condition is also necessary, because all pure
two-particle entangled states can be made to violate such a Bell
inequality by an appropriate choice of the observables
\cite{GISIN2,POPROHR}.

For $N$-particle systems, generalized Bell inequalities have been
reported by Mermin \cite{MERMIN} and Ardehali \cite{ARDEHALI}.
These $N$-particle inequalities are likewise derived under the
assumption of local realism. More explicitly, it is assumed that
each particle may be assigned independent elements of reality
corresponding to  certain measurement outcomes. A bound on the
expected correlations is then obtained and shown to be violated by
the corresponding quantum mechanical expectation values by a
maximal factor that grows exponentially with $N$
\cite{MERMIN,ARDEHALI}. $N$-particle experiments that violate
these inequalities are then, again, disproofs of the assumptions
of local realism.

However,  the violation of local realism is not sufficient for
confirmation of the entanglement of all $N$ particles. For this
purpose, one must also address the question of whether the data
admit a model in which less than $N$ particles are entangled.  The
standard generalized Bell inequalities mentioned above are not
designed to deal with this issue, and thus, leave the loophole
open that the data might be explained by mixtures of states in
which less than $N$ particles are entangled.
 In fact, as shown in more detail below, the data of some experiments
 aimed to produce three-particle entanglement
\forget{for the experiment by Bouwmeester \emph{et
al.}~\cite{BOU1998},} may be approximated surprisingly closely by
a mixture of two-particle entangled states. This forms the
motivation for a closer investigation of conditions needed to
close this particular loophole.

Some conditions of this kind have been formulated in the recent
literature \cite{LEWENSTEIN} in terms of partial transpositions of
the $N$-particle density matrix.
 Unfortunately, it is not clear at
present how these conditions may be tested experimentally.
 In this paper, we review two
experimentally accessible conditions, presented in Sec.~2 as
conditions \emph{A} and \emph{B}. In Sec.~3, we  analyze some
recent experiments \cite{BOU1998,RAUSCH,PAN} to produce
three-particle entanglement, in order to see whether they meet
these conditions.
%%%%%%%%%%%%%%%%%%%%%%%%%%%%%%%%%%%%%%%%%%%%%%%%%%%%%%%%%%%%%%%%%%%%%%%%%%%%%%%%%%
\forget{. We would like to stress that these
    conditions are experimentally accessible, whereas other,
    perhaps more stringent, conditions for entanglement
    \cite{LEWENSTEIN} using the partial transposition criterium
    and properties of positive maps are not
     experimentally accessible.}%
\forget{
    We would like to stress that these conditions
    are experimentally accessible, this in contradistinction to other,
    perhaps more stringent, existing conditions for entanglement such as using the partial
    transposition criterium \cite{LEWENSTEIN}, which are however not
    experimentally accessible.}%
\forget{ We would like to stress that these conditions
    are experimentally accessible, this in contradistinction to other existing
    conditions for entanglement using the partial
    transposition criterium and properties of positive maps \cite{LEWENSTEIN},
    although perhaps more stringent,which are not
    experimentally accessible. }%
\forget{
    It is the purpose of this letter to point out a loophole problem
    in the data analysis of recent experiments by Bouwmeester \emph{et
    al.}~\cite{BOU1998}, by Pan \emph{et al.}~\cite{PAN} and of
    Rauschenbeutel \emph{et al.}~\cite{RAUSCH} which claim to have
    succeeded in creating and detecting true three-particle entangled
    states. The loophole is that, although at first indicating three
    particle entanglement, the experimental data might in fact be
    reproduced by only two-particle entangled states. This loophole is
    hinted at by the analysis presented in Appendix A which shows that
    most of the salient results of the experiment by Bouwmeester
    \emph{et al.}~\cite{BOU1998} can be approximated surprisingly
    closely by means of a mixture of two-particle entangled states.
    This (possible loophole) forms a really strong motivation for a
    closer investigation of the subject which is presented in this
    letter.

    In trying to settle the problem we have analysed the three
    experiments to see whether or not they meet any of the above
    mentioned sufficient conditions \emph{A} and \emph{B} for true
    three-particle entanglement ($N$=$3$). }%%%%%%%%%%%%%%%%%%%%%%%
%%%%%%%%%%%%%%%%%%%%%%%%%%%%%%%%%%%%%%%%%%%%%%%%%%%%%%%%%%%%%%%
It is shown that this is not the case. This, of course, does not
prove that there is no three-particle entanglement in these
experiments. Rather, we conclude that on the basis of the
conditions reviewed here, the above loophole problem remains
unresolved.
  However, we propose modifications of the
experimental procedure that would allow for a more definite
confirmation of three-particle entanglement. \forget{
    To be sure, we do not claim that in the
    experiments here considered there is no
    three-particle entanglement, but merely that pointing out the loophole problem
    leaves the matter as yet unresolved.}

\section{Sufficient conditions for N-particle entanglement}
We start with the definition of the basic concept.
 Consider an
arbitrary  $N$-particles system described by a Hilbert space
${\cal H} = {\cal H}_1 \otimes \cdots \otimes {\cal H}_N$. A
general mixed state $\rho$ of this system is called $N$-particles
entangled  iff no convex decomposition of the form
\begin{equation}
\rho = \sum_{i} p_i   \rho_i,
~~~\mbox{ with $p_i\geq 0$, $\sum_i p_i =1$,}
\label{factorisability0}
\end{equation}
exists
 in which  all the states $\rho_i$  are
factorizable into products of states of less than $N$ particles.
Of course, since each factorizable mixed state is a mixture of
factorizable pure states, one may equivalently assume that
factorizable states $\rho_i$ are pure, so that the decomposition
(\ref{factorisability0}) takes the form
\begin{equation}
\rho = \sum_{i} p_i\ket{\psi_i}\bra{\psi_i}.
\label{factorisability}
\end{equation}

In order to extend  the above terminology,
  let $K$ be  any subset $K \subset \{1, \ldots, N\}$
   and let $\rho^K$ denote a state of the subsystem
composed of the particles labeled by $K$.
 We will call an $N$-particle state  $M$ particle
 entangled ($M<N$)
 iff a decomposition exists of the form
\begin{equation} \rho = \sum_{i}
p_i\, \rho_i^{ K^{(i)}_1}\otimes \cdots \otimes
\rho_{i}^{K_{r_{i}}^{(i)}} \label{factorisability2}
\end{equation}
where, for each $i$,  $K^{(i)}_1, \ldots, K^{(i)}_{r_i}$ is some
partition of $\{ 1, \ldots ,N\}$ into $r_i$ disjoint subsets, each
subset  $K^{(i)}_j$ containing at most $M$ elements; but no such
decomposition is possible when these subsets are required to
contain less than $M$ elements.

An example of an $N$-particle state that is $N$ particle entangled
is the Greenberger-Horne-Zeilinger (GHZ) state
\begin{equation}
\ket{\psi_{\rm GHZ}}=
\frac{1}{\sqrt{2}}\left(\ket{\uparrow\uparrow\cdots\uparrow}
+\ket{\downarrow\downarrow\cdots\downarrow}\right),
\label{Npartstate}
\end{equation}
where  $\ket{\uparrow}$ and $\ket{\downarrow}$ denote the
eigenstates of  some dichotomic observable (e.g.\ spin or
polarization) which we will take, by convention, as oriented along
the $z$ axis.  On the other hand, the  three-particle state
\begin{equation} \rho =  \frac{1}{2}(
 \op{P}^{(1)}_{\uparrow}\otimes \op{P}^{(23)}_S +
\op{P}^{(1)}_{\downarrow}\otimes \op{P}^{(23)}_T)
 \label{mix}\end{equation} is only two-particle entangled.
 Here,
$\op{P}^{(23)}_T$ and $\op{P}^{(23)}_S$ denote projectors on the
triplet state $ \frac{1}{\sqrt{2}}(
\ket{\uparrow\downarrow}+\ket{\downarrow\uparrow})$ and singlet
state $ \frac{1}{\sqrt{2}}(\ket{\uparrow
\downarrow}-\ket{\downarrow \uparrow})$, respectively,  for
particles 2 and 3, and
$\op{P}{(1)}_{\downarrow}=\ket{\downarrow}\bra{\downarrow}$ and
$\op{P}^{(1)}_{\uparrow}=\ket{\uparrow}\bra{\uparrow}$ are the
``down'' and ``up'' states for particle 1. Note that, as the state
(\ref{mix}) exemplifies, an $N$-particle state can be $M$ particle
entangled even if it has no $M$-particle subsystem whose (reduced)
state is $M$ particles entangled. In the remainder of this
section, we review two inequalities that allow for a test between
$N$-particle and $M$-particle entangled states, focusing mainly on
$N=3$ and $M=2$.

\textbf{Condition A}:  The following  condition  has been derived
by Gisin and Bechmann-Pasqui\-nucci \cite{GISIN} for a system of
$N$ two-level particles (q-bits).  As a start, consider the
well-known Bell-CHSH inequality \cite{CHSH} for two particles. Let
$A$ and $A'$ be dichotomous observables on the first particle,
with possible outcomes $\pm 1$, and similarly  for observables $B$
and $B'$ on the second particle. Consider the expression
\begin{equation}
F_2 := AB +A'B+AB'-A'B'=(A+A')B +(A-A')B' \leq 2.
\label{2deeltjes}
\end{equation}
 Assuming local realism, the pair  $A$  and $B$
 are conditionally  independent
\be p^{\rm lr}_{AB} (a,b) = \int_\Lambda p_A(a|\lambda)
p_B(b|\lambda) \rho(\lambda) d\lambda  \ee  and similarly for the
pairs $A',B$, $A,B'$, and $A',B'$, where $p_{A}$ and $p_B$ are
probabilities conditional on the hidden variable $\lambda \in
\Lambda$. If we denote the expected correlations as
\[ E_{\rm lr}(AB) =   \sum_{ab} ab \, p^{\rm lr}_{AB}(a,b),  \]
 we obtain the standard
 two-particle Bell-CHSH inequality \cite{CHSH}:
  \be |E_{\rm lr} (F_2)| = \left|(E_{\rm lr}(AB)+ E_{\rm
lr}(A'B)+ E_{\rm lr}(AB')- E_{\rm lr}(A'B')\right| \leq 2.\ee

 In
quantum mechanics the observable $A$ is represented by the spin
operator $\op{A} =\mathbf{\vec{a}\cdot\vec{\sigma}}$ with unit
three-dimensional vector $\mathbf{\vec{a}}$, and similarly for the
other three observables. The expected correlation in a state
$\rho$ is given by $E_\rho(AB)= \Tr ( \rho\, \mathbf{\vec{a}\cdot
\vec{\mathbf{\sigma}}\otimes \mathbf{\vec{b}\cdot
\vec{\sigma}}})$.
   In terms  of these expectation values the
   Bell-CHSH inequality  may be violated by entangled quantum states.
The largest violation of this inequality by a quantum state is
$2\sqrt{2}$ \cite{CIRELSON}.

The Bell-CHSH inequality is generalized by Gisin and
Bechmann-Pasquinucci  to $N$ particles through a recursive
definition. Let  $A_j$
  and $A'_j$  denote dichotomous observables on the $j$th particle, ($j= 1,2,\ldots, N$), and define
  \begin{equation}
  F_N := \frac{1}{2}( A_N+A_N')F_{N-1} +\frac{1}{2}(A_N-A_N')F_{N-1}' \leq
  2,
  \end{equation}
where $F_{N-1}'$ is the same expression as $F_{N-1}$ but with all
$A_j$ and $A_j'$ interchanged.  Here, the upper bound on $F_N$
follows by natural induction from the bound (\ref{2deeltjes}) on
$F_2$. One now obtains
  the so-called Bell-Klyshko
  inequality \cite{GISIN},
\begin{equation}
   | E_{\rm lr} (F_N )| \leq 2.
\label{Npart}
\end{equation}
This Bell-Klyshko inequality is also violated in quantum
mechanics. That is to say, the expectation value of the
corresponding operator
\begin{equation}
  \op{F}_N := \frac{1}{2}( \op{A}_N+\op{A'}_N)\otimes\op{F}_{N-1} +\frac{1}{2}(\op{A}_N-\op{A'}_N)\otimes\op{F'}_{N-1} \leq 2
  \end{equation}
may violate the bound (\ref{Npart}) for entangled quantum states.
As shown in reference \cite{GISIN}, the maximal value is \be
|E_\rho (\op{F}_N)| \leq 2^{(N+1)/2} \label{maxN}, \ee i.e., a
violation by a factor $2^{(N-1)/2}$.

The inequality (\ref{Npart}) may now be extended into a test of
$N-1$-particle entanglement. Consider a  state in which one
particle (say the $N$th) is independent  from the others, i.e.:
$\rho = \rho_{\{N\}} \otimes \rho_{\{1,\ldots, N-1\}}$.
 One then obtains
 \BE
 |E_\rho(\op{F}_N) | &=&
   \left| \Tr \rho\left( \frac{1}{2} \bigl( \op{A}_N  +
\op{A'}_N\bigr)\otimes \op{F}_{N-1}
 +  \frac{1}{2}\bigl( \op{A}_N  - \op{A'}_N\bigr)\otimes \op{F'}_{N-1} \right)  \right|  \nn\\
   & =&
\frac{1}{2} \left|\left( \langle \op{A}_N \rangle_{\rho} + \langle
\op{A'}_N \rangle_{\rho_{}} \right)\,
 \Tr \rho \op{F}_{N-1}
 + \bigl( \langle
 \op{A}_N\rangle_\rho  -
\langle \op{A'}_N \rangle_\rho \bigr) \, \Tr  \rho \op{F'}_{N-1}
\right|
     \nn\\
&=&
 \frac{1}{2} \left| \langle \op{A}_N \rangle_{\rho}\, \left(( E_{\rho_{}}
(\op{F}_{N-1}) + E_{\rho_{}} (\op{F'}_{N-1}) \right) + \langle
\op{A'}_N\rangle_{\rho_{}} \,\left( E_{\rho_{}}(\op{F}_{N-1})
-E_{\rho_{}}(\op{F'}_{N-1})\right)
 \right|\nn\\
&\leq&    \frac{1}{2}|   E_\rho( \op{F}_{N-1}) + E_\rho(
\op{F'}_{N-1}) | +
 \frac{1}{2}|E_\rho( \op{F}_{N-1}) - E_\rho( \op{F'}_{N-1})  | \nn\\
 &= & \max ( |E_\rho(\op{F}_{N-1})|,  |E_\rho(\op{F'}_{N-1})|) \leq 2^{N/2} \EE
where we have used $|\langle \op{A}_N\rangle|\leq 1, |\langle
\op{A'}_N\rangle|\leq 1$ and the bound (\ref{maxN}).

Since ${\op{F}}_N$ is invariant under a permutation of the $N$
particles, this bound holds also for a  state in which another
particle than the $N$th factorizes, and, since $E_\rho(F_N)$ is
convex as a function of $\rho$, it holds also for
 mixtures of such  states. Hence, for every $(N-1)$-particle
entangled state we have
 \be |E_\rho(\op{F}_N)| \leq 2^{N/2}. \label{v}
\ee
 Thus, a  sufficient
condition for  $N$-particle entanglement is a violation of
Eq.~(\ref{v}), i.e.,  inequality (\ref{Npart}) should be violated
by a factor larger than $2^{(N/2-1)}$.

 Specializing now to the case where $N=3$, inequality (\ref{v})
 may be written more conveniently as \be
\left|E_{}(ABC')+ E_{}(AB'C)+ E_{}(A'BC)- E_{\rm}(A'B'C')\right|
\leq 2^{3/2}, \label{3deeltjes} \ee where we have put $A_1 = A ,
A_2=B$, and $A_3= C$.

For example, for a choice of spin directions $\vec{a}=\vec{a'}$
along the $z$ axis, and $\vec{b}$, $\vec{b'}$, $\vec{c}$,
$\vec{c'}$ in the $xy$ plane with angles $\beta =0 ,~ \beta' =
\pi/2,~
 \gamma=  \pi/4$, and  $\gamma'= -\pi/4$ from the $x$axis,
 the mixed state (\ref{mix}) gives $E_{\rho}(F_3)= 2\sqrt{2}$.
 This violates
inequality (\ref{Npart}), thus indicating
 two-particle entanglement,  but does  not violate inequality (\ref{3deeltjes}),
 and thus shows no three-particle entanglement.

 \textbf{\textit{Condition B}}:
Another condition for $N$-particle entanglement follows from the
fact that the internal correlations of a quantum state are encoded
in the off-diagonal elements of the density matrix that represents
the state in a product basis. We summarize here  the derivation
presented by Sackett \emph{et al.}~\cite{SACKETT}. Consider the
so-called state preparation fidelity $F$ of a $N$-particle state
$\rho$ defined as
\begin{equation}
 F(\rho) := \langle \psi_{\rm GHZ} |\rho|\psi_{\rm GHZ}\rangle= \frac{1}{2} (P_{\uparrow}+P_{\downarrow})
 +\mbox{Re~} \rho_{\uparrow \downarrow},
 \label{fidelity}
\end{equation}
where $\ket{\psi_{\rm GHZ}}$ is given by (\ref{Npartstate}),
$P_{\uparrow} :=\bra{\up\cdots \up} \rho\ket{\up \cdots\up}$,
$P_{\downarrow}:=
\bra{\downarrow\downarrow\cdots\downarrow}\rho\ket{\downarrow\downarrow\cdots\downarrow}$
  and $\rho_{\up
\down} :=
\bra{\up\up\cdots\up}\rho\ket{\downarrow\downarrow\cdots\downarrow}$
is the far off-diagonal matrix element in the  $z$ basis.
 Now partition the set of $N$ particles into
two disjoint subsets $K$ and $K'$
   and consider a pure state of the form
  \begin{equation}
  \ket{\phi}=
  \left(
   a\ket{\uparrow\uparrow\cdots\uparrow}^K
  + \cdots +b\ket{\downarrow\downarrow\cdots\downarrow}^K  \right)
  \otimes
  \left(c\ket{\uparrow\uparrow\cdots\uparrow}^{K'} +\cdots
  +d\ket{\downarrow\downarrow\cdots\downarrow}^{K'}
\right),
  \label{Sackett1}
  \end{equation}
where $\ket{\uparrow\uparrow\cdots\uparrow}^K $ is the state with
all particles in subset $K$ in the ``up''-state and similarly for
the other terms. Normalization of $\ket{\phi}$ leads to $|a|^2
+|b|^2\leq1$ and $|c|^2+|d|^2\leq1$. It then follows that
  \begin{equation} 2F( \ket{\phi}\bra{\phi}) =
  |ac|^2 + |bd|^2  + 2\,\mbox{Re~}( ab^* cd^*)  \leq
   \left((|a|^2 + |b|^2\right) \left(|c|^2 +|d|^2\right) \leq 1.
  \label{Sackett3}
  \end{equation}
  Thus, the state preparation
fidelity is at most  $1/2$ for any  state of the form
(\ref{Sackett1}). From the convexity of $F(\rho)$ it follows that
this inequality also holds for any mixture of such  product
states, i.e.\ for any state
 $\rho$ as defined in
Eq.~(\ref{factorisability}).

We have thus found a second sufficient condition for $N$-particle
entanglement, namely,
 \be F(\rho) > 1/2. \ee
 Of course, analogous conditions may be obtained by replacing the
special state $\ket{\psi_{\rm GHZ}}$ in definition
(\ref{fidelity}) by another maximally entangled state, such as
$\frac{1}{\sqrt{2}}(\ket{\up\ldots\up\down} \linebreak[1] \pm
\ket{\down\ldots \down \up})$, etc.
 An experimental test of condition \emph{B} requires the
 determination of
the real part of the far off-diagonal matrix element
$\rho_{\uparrow \downarrow}$. Now, obviously, $\mbox{Re~}
\rho_{\up\down}$ is not the  expectation value of a product
observable, and information about this quantity  may only be
obtained indirectly. In the next section we discuss several
experimental procedures by which this information may be obtained.
As we shall see,  it is important that such procedures make sure
that no unwanted matrix elements contribute to the determination
of this quantity.

\section{Analysis of experiments}
 Using the conditions \emph{A} and \emph{B} discussed above, we now turn
to the analysis of three recent experimental tests for
three-particle entangled states.

(\textbf{I}). In the experiment of Bouwmeester \emph{et
al.}~\cite{BOU1998}, the three-photon entangled state
$\ket{\psi_{\rm B}}= \frac{1}{\sqrt{2}}\left( \ket{HHV} +\ket{VVH}
\right)$ is claimed to be experimentally observed. Here, $\ket{H}$
and $\ket{V}$ are the horizontal and vertical polarization states
of the photons. We represent  this state in the $z$ basis using
$\ket{H}=\ket{\uparrow}$ and $\ket{V}=\ket{\downarrow}$ as \be
\label{B} \ket{\psi_{\rm
B}}=\frac{1}{\sqrt{2}}\left(\ket{\uparrow\uparrow\downarrow}
+\ket{\downarrow\downarrow\uparrow}\right).\ee  The experiment
consisted, first, of a set of threefold coincidence measurements
in the $zzz$ directions, in which the fraction of the desired
outcomes, i.e., the components $\ket{\up\up\down}$ and
$\ket{\down\down \up}$ out of the $2^3$ possible outcomes was
determined and found to be in a ratio of 12:1. Furthermore, to
show coherent superposition of these components a second set of
measurements was performed in the $xxx$ directions. For a large
fraction of the observed data,  this second set of measurements
shows correlations as expected from the desired state
$\ket{\psi_B}$. A third series of measurements performed in the
$zxx$ directions showed no such correlations, again, as expected
from the state $\ket{\psi_{\rm B}}$. Bouwmeester \emph{et al.}
concluded that: ``The data clearly indicate the absence of
two-photon correlations and thereby confirm our claim of the
observation of GHZ entanglement between three spatially separated
photons \cite{BOU1998}.'' However, no quantitative analysis was
made to determine whether two-particle entangled states may
account for or contribute to the observed data. In order to show
that such an analysis is not superfluous, it is shown in Appendix
A how most of the salient results of this experiment may in fact
be reproduced by a simple two-particle entangled state.
 Thus, we are presented with the loophole problem whether or not the
observed data may be regarded as hard evidence for
  true three-particle entanglement.

The experiment of Pan \emph{et al.}~\cite{PAN}, performed by the
same group, aimed to produce the GHZ state $\ket{\psi_{\rm
GHZ}}=\frac{1}{\sqrt{2}}\left(\ket{\uparrow\uparrow\uparrow}
+\ket{\downarrow\downarrow\downarrow}\right)$ by a procedure
similar to the previous experiment.
  Although their main goal was to
show  a conflict with local realism, Pan \emph{et al.} also claim
to have provided evidence for three-particle entanglement.  For
this purpose, they performed  four series of measurements, in the
$xxx, xyy, yxy$, and $yyx$ directions, and tested a three-particle
Bell inequality of the form derived by Mermin~\cite{MERMIN}. This
inequality is presented in \cite{BOU2000} and reads \be|\langle
xyy\rangle+ \langle yxy\rangle + \langle yyx \rangle- \langle xxx
\rangle| \leq 2, \label{mermin} \ee where $\langle xyy\rangle$ is
the expectation value of $\sigma^{(1)}_x \otimes\sigma^{(2)}_y
\otimes \sigma^{(3)}_y$, etc. The reported experimental data are
 \be|\langle xyy\rangle+ \langle yxy\rangle
+ \langle yyx \rangle- \langle xxx \rangle|=2.83 \pm 0.09,
\label{result} \ee in clear violation of Eq.~(\ref{mermin}).
However, as mentioned in the Introduction, violating a
generalized Bell inequality of this type is not sufficient to
confirm three-particle entanglement.
 Thus, again, the question remains
whether the reported data may be regarded as confirmation of
three-particle entanglement. In particular, one might ask,  do
these experiments meet either of the conditions \textit{A} or
\textit{B}?

Upon further analysis, we may answer this question. First, we note
that the procedure followed by Bouwmeester \emph{et al.} does not
allow for a test of condition \emph{A} even in the ideal case
where
 the desired state is actually produced.
This is because measurements were performed only in various
directions in the $xz$ plane.  However, for any observable
$\mathbf{\vec{a} \cdot \vec{\sigma} \otimes \vec{b}\cdot\vec{
\sigma}\otimes \vec{c} \cdot \vec{\sigma}}$ with $\mathbf{\vec{a},
\vec{b}, \vec{c}} $ unit vectors
 in the $xz$ plane, we obtain $\bra{\psi_{\rm B}}\mathbf{\vec{a} \cdot
\vec{\sigma} \otimes \vec{b}\cdot\vec{ \sigma}\otimes \vec{c}
\cdot \vec{\sigma}}\ket{\psi_{\rm B}} = \cos\alpha \cos\beta
\cos\gamma$, with $\alpha$, $\beta$, and $\gamma$ the angles these
vectors span from the $x$ axis. These expectation values are
 factorizable, and  measurements of spin observables in the $xz$ plane cannot
 lead to a violation of condition \emph{A}, i.e., the
inequality (\ref{3deeltjes}). Neither does the choice of
measurements in this experiment allow for a test of condition
\emph{B}.\@ For such a test, one would have to determine the
relevant state preparation fidelity, i.e., $\bra{\psi_{\rm B}}
\rho \ket{\psi_{\rm B}}$ of the experimentally produced state
$\rho$. But the reported data do not allow for an estimate of the
relevant off-diagonal element $\mbox{Re~} \bra{\up\up\down}\rho
\ket{\down\down \up}$. Indeed, the only measurements that are
sensitive to the value of this matrix element, namely, those in
the $xxx$ directions, are also sensitive to all other matrix
elements on the cross diagonal in the $zzz$ eigenbasis.

The experiment by Pan \emph{et al.} is more rewarding in this
respect.
 The inequality (\ref{mermin}) tested in this experiment is identical to
 a Bell-Klyshko inequality~(\ref{Npart}) for  $N=3$.
 Since the inequality is violated, the experiment
is indeed a violation of local realism. However, within
experimental errors, the measured value $E(F_{3}) = 2.83 \approx
2^{3/2} $ does not violate inequality (\ref{3deeltjes})
 that would be sufficient for evidence of
three-particle entanglement. Thus, although the experimental
procedure  allowed for a test of Condition \emph{A}, it did not
violate it.
  Further, the experiment
of Pan \emph{et al.} did not attempt to test condition \textit{B}
either.

However, both experiments may be simply adjusted to test both
conditions. \forget{Here we report results obtained by Popescu
\cite{POPESCU2}, who has specified the appropriate basis and
angles to measure in order for these experiments to test condition
\emph{B}.}  If, in the experiment of Bouwmeester \emph{et al.},
one measures spin observables in directions $\mathbf{\vec{a}}$,
$\mathbf{\vec{b}}$ and $\mathbf{\vec{c}}$  in the $xy$ plane,
rather than the $xz$ plane, one obtains $E_{\ket{\psi_{\rm
B}}}(ABC)= \bra{\psi_{\rm B}}
\mathbf{\vec{a}\cdot\mathbf{\vec{\sigma}}}\otimes
\mathbf{\vec{b}\cdot\vec{\sigma}} \otimes
\mathbf{\vec{c}\cdot\vec{\sigma}} \ket{\psi_{\rm B}}= \cos (\alpha
+ \beta - \gamma)$
 where  $\alpha$,
$\beta$, and $\gamma$  again denote the angles from the $x$ axis.
For the choice: $\alpha= \pi/2,~\alpha'=0,~\beta=
\pi/4,~\beta'=-\pi/4,~\gamma=\pi/4$, and $\gamma'=3\pi/4$,
 the inequality (\ref{3deeltjes}) will be violated maximally by the
value $4$. \forget{ Similarly, for $\alpha=
0,~\alpha'=-\pi/2,~\beta= \pi/4,~\beta'=-\pi/4,~\gamma=0$, and
$\gamma'=\pi/2$, the inequality (\ref{SVETCH2}) will be violated
maximally by the value $4 \sqrt 2$.}

For the state $\ket{\psi_{\rm
GHZ}}=\frac{1}{\sqrt{2}}(\ket{\uparrow\uparrow\uparrow}+\ket{\downarrow\downarrow\downarrow})$,
used in the experiment of Pan \emph{et al.}, it follows likewise
that $E_{\rm GHZ}(ABC)= {\bra{\psi_{\rm GHZ}}
\mathbf{\vec{a}\cdot\mathbf{\vec{\sigma}}}\otimes
\mathbf{\vec{b}\cdot\vec{\sigma}} \otimes
\mathbf{\vec{c}\cdot\vec{\sigma}} \ket{\psi_{\rm GHZ}}}= \cos
(\alpha + \beta + \gamma)$ when the vectors are chosen in the
$xy$ plane. Then, inequality (\ref{3deeltjes})  will be violated
maximally by the value $4$ for the choice: $\alpha=
\pi/2,~\alpha'=0,~\beta= \pi/2,~\beta'=0,~\gamma=\pi/2$, and
$\gamma'=0$. \forget{ The inequality (\ref{SVETCH2}) is violated
by the maximum value $4 \sqrt 2$ for the angles $\alpha=
0,~\alpha'=-\pi/2,~\beta= \pi/4,~\beta'=-\pi/4,~\gamma=0$, and
$\gamma'=-\pi/2$.}
 Using these angles in
future experiments will thus allow for tests of three-particle
entanglement.

Finally, we  discuss how the experiments can be adjusted
 in order to test  condition \emph{B}.
 Determining the populations $P_\up$ and $P_\down$ in Eq.~(\ref{fidelity}) is rather trivial and
will not be discussed. Here, we mention two possible procedures to
determine  $\mbox{Re~} \rho_{\uparrow \downarrow}$.
 The first is to use a three-particle analogue of the
method used by Sackett \emph{et al.}~\cite{SACKETT}. Consider the
observable $\op{S}_{\pm}(\phi):= \mathbf{\vec{n}_\phi
\cdot\vec{\sigma}\otimes \vec{n}_\phi \cdot\vec{\sigma}\otimes
 \vec{n}_{\pm \phi}\cdot \vec{\sigma}}$ where
 $ \mathbf{\vec{n}}_\phi =
(\cos\phi, \sin \phi, 0)$. The expectation values $\bra{\psi_{\rm
GHZ}}\op{S}_+ (\phi)\ket{\psi_{\rm GHZ}}$ and $\bra{\psi_{\rm
B}}\op{S}_-(\phi)\ket{\psi_{\rm B}}$, considered as functions of
$\phi$, oscillate as $A\cos (3\phi +\alpha_0) + B\cos( \phi
+\beta) + \mbox{const.}$, where $ A = 2\,\mbox{Re~}
\rho_{\up\down}$. (That is, $A = 2\,\mbox{Re~}
\bra{\up\up\up}\rho\ket{\down \down \down}$ in the first, and $A
=2\,\mbox{Re~} \bra{\up\up\down}\rho\ket{\down \down \up}$ in the
second case.)
 Hence, by measuring $S_+(\phi)$ for the GHZ state
(\ref{Npartstate}), or $S_-(\phi)$ for the state (\ref{B}),
 for a variety of angles $\phi$, and by filtering
out the amplitude that oscillates as $\cos 3\phi$, one obtains an
estimate of the relevant off-diagonal element $| \mbox{Re~}
\rho_{\up\down}|$ needed to test Condition \emph{B}.

However, a simpler way to determine this  off-diagonal matrix
element is to take advantage of the simple  operator identity: \be
 \sigma_x \otimes \sigma_y \otimes \sigma_y +
 \sigma_y\otimes \sigma_x \otimes \sigma_y  +
 \sigma_y \otimes \sigma_y \otimes \sigma_x -
 \sigma_x \otimes \sigma_x \otimes \sigma_x
 =  - 4\left( \ket{\down\down\down}\bra{\up\up\up} +
 \ket{\up\up\up}\bra{\down\down\down}\right),
 \ee
 so that for all states $\rho$
 \be \langle x y y +
 yxy  +
 y  y x -
 x x x \rangle_\rho
 =
  - 8 \,\mbox{Re~} \bra{\up\up\up}\rho
 \ket{\down\down\down}. \label{usef}\ee
 Since the expectation value in the left-hand side of Eq.~(\ref{usef})
has already been measured in the experiment of Pan \emph{et al.},
one may
 infer  from their reported result (\ref{result})
 that
\[   |\mbox{Re~}(\rho_{\up\down})| =\frac{2.83 \pm 0.09}{8} =0.35  \pm 0.01 .\]
Thus, only one additional measurement in the $zzz$ directions
would have been sufficient for a full test of condition \emph{B}.
If the ratio reported in the experiment of Bouwmeester \emph{et
al.} of 12:1 (corresponding to populations of 0.40) is a feasible
result for the setup of Pan \emph{et al.} too, one should expect
to obtain an experimental value of $F(\rho) \approx 0.75$, well
above the threshold value of $1/2$.

(\textbf{II}).  The experiment of Rauschenbeutel \emph{et
al.}~\cite{RAUSCH} was set up to measure three-particle
entanglement for three spin-$\frac{1}{2}$ systems (two atoms and a
single-photon  cavity field mode). The state of the cavity field
is not directly observable, and was therefore copied onto a third
atom, so that the actual measurement was carried out on a
three-atom system.
 Let us first adapt the notation of
\cite{RAUSCH} to the notation of this paper: Their target
three-atom state $\ket{\Psi_{\rm triplet}} =
\frac{1}{\sqrt{2}}(\ket{{e}_1,{i}_2,{g}_3} + \ket{{g}_1,
{g}_2,{e}_3})$
 is  represented here as $\ket{\psi_{B}}=
\frac{1}{\sqrt{2}} ( \ket{\uparrow\uparrow\downarrow} +
       \ket{\downarrow\downarrow\uparrow})$.

Condition \emph{B} was used to test for three-particle
entanglement. The measured fidelity is claimed to be $F=0.54 \pm
0.03$ and this is, within experimental accuracy, only just
greater than the sufficient value of 1/2. However, we will argue
that upon a ``worst-case'' analysis of the data this result may no
longer be claimed to hold, since one cannot exclude that other
off-diagonal density-matrix elements contribute to their
determination of $\mbox{Re~} \rho_{\uparrow \downarrow}$.

 In the experiment, first the individual populations of eigenstates in the
$zzz$ directions was determined. These populations are the
so-called longitudinal correlations in Fig. 3 of \cite{RAUSCH} and
give the following results: (all numbers $\pm 0.01$) \be
 \begin{array}{|c|c|c|c|c|c|c|c|} \hline
  P_{\up\up\up} & P_{\up\up\down} & P_{\up\down\up}
&P_{\up\down\down} & P_{\down\up\up} & P_{\down\up\down} &
P_{\down\down\up} &P_{\down\down\down}\\ \hline
  0.1  & 0.22 & 0.06& 0.04 & 0.1 & 0.09 & 0.36 & 0.03 \\ \hline
\end{array} \label{tab}
\ee This gives $ \frac{1}{2}(P_{\up\up\down} +P_{\down\down\up})=
0.29$.
 Next, the off-diagonal matrix element
$\mbox{Re~} \bra{\up\up\down}\rho\ket{\up\up\down}$ is determined
by first projecting particle $2$ onto either $\ket{+}_x$ or
$\ket{-}_x$, and  measuring the so-called `Bell signals' $
\op{B}_{\pm}(\phi) := \sigma^{(1)}_{x}\otimes \mathbf{
\vec{n}}_\phi \cdot \sigma^{(3)}$ on the remaining pair. Here,
again, $ \mathbf{\vec{n}}_\phi = (\cos\phi, \sin \phi, 0)$.

Thus, the expectation of these Bell signals is given by $ \langle
\op{B}_{\pm}(\phi)\rangle = \Tr(\rho\, \sigma^{(1)}_{x} \otimes
\op{P}^{(2)}_{\pm}\otimes \mathbf{\vec{n}}_\phi \cdot
\vec{\sigma}^{(3)})$.
 The Bell signal $\langle \op{B}_{+}(\phi)\rangle$ is predicted to oscillate as
$A \cos \phi$. The other Bell signal
$\langle\op{B}_-(\phi)\rangle$ has a phase shift of $\pi$ and thus
oscillates as $-A\cos \phi$. In the case of the desired
three-particle state (\ref{B}),
 the amplitude $A$ of the oscillatory Bell signals is
equal to $A= 2|\bra{\up\up\down}\rho\ket{\down \down \up}|$.
  The experimental data
give a value of $A = 0.28 \pm 0.04$, leading to the result $F=
\frac{1}{2}(P_{\up\up\down} + P_{\down\down\up} +A) =0.54 \pm
0.03$.

 However, if one assumes a general unknown state, it turns out
that not only the matrix element $\bra{\up\up\down}\rho\ket{\down
\down \up}$ (and its complex conjugate), but also the elements
$\bra{\up\up\up}\rho\ket{\down \down \down}$,
$\bra{\up\down\down}\rho\ket{\down \up\up}$
  and
$\bra{\up\down\up}\rho\ket{\down  \up\down }$ and their respective
complex conjugates contribute to the measured amplitude $A$.
 In a ``worst-case''
analysis, these unwanted density matrix elements should be
assigned the highest possible value compatible with the values of
the measured populations in table (\ref{tab}). Suppose these
contributions sum up to the maximal value $w$ in the amplitude
$A$, then we may conclude that $2\, \mbox{Re~} \rho_{\uparrow
\downarrow}$ has the ``worst-case'' value of $A-w$.

Using the data from \cite{RAUSCH}, such an analysis has been
performed from which we obtain $w= 0.26  \pm0.04$  (see Appendix B
for details). $2\, \mbox{Re~}\rho_{\uparrow \downarrow}$ then has
the approximate value of $0.02 \pm 0.05$ instead of the value
$0.28 \pm 0.04$ reported by Rauschenbeutel \emph{et al. } This
value gives an approximate fidelity $F=0.31 \pm 0.05$, which no
longer meets the inequality $F\geq 1/2$ of    Condition \emph{B}.

One might object to our worst case analysis because it assumes a
maximal contribution from other three-particle entangled states.
This is not only physically implausible, but would also give rise
to the hope that at least \emph{some} three-particle entangled
state has been observed. The prospects of this hope are difficult
to assess. Of course, one has to take  into account that a mixture
of different three-particle entangled states is not necessarily a
three-particle entangled state. But it is difficult to say whether
or not this holds for the worst-case mixture discussed in Appendix
B.

However this may be, it is straightforward to show that the
unwanted matrix elements may contaminate the data from this
experiment even for two-particle entangled states. For example,
consider  the incoherent mixture of two pure Bell signal states
 \begin{equation} \rho_{\rm mix} =
\frac{1}{2}\left(
 \op{P}^{(2)}_{+}\otimes \op{P}^{(13)}_S +
\op{P}^{(2)}_{-}\otimes \op{P}^{(13)}_T \right),
 \label{mix'}\end{equation}
 where
$\op{P}^{(13)}_T$ and $\op{P}^{(13)}_S$ denote projectors on the
triplet state $ \frac{1}{\sqrt{2}}(
\ket{\uparrow\downarrow}+\ket{\downarrow\uparrow})$ and singlet
state $ \frac{1}{\sqrt{2}}(\ket{\uparrow
\downarrow}-\ket{\downarrow \uparrow})$ respectively  for the
particles 1 and 3, and $\op{P}^{(2)}_{\pm}$ are the
eigenprojectors in the $x$ direction for particle 2.  For this
state, the expected values  of $P_{\up\up\down}$ and
$P_{\down\down\up}$ are 0.25, and $A := \max_\phi |\Tr \rho_{\rm
mix} \op{B}_{\pm}(\phi) | = 1$, while $\bra{\up\up\down}\rho_{\rm
mix} \ket{\down\down \up} = \frac{1}{4}$. In the experimental
procedure of Rauschenbeutel \emph{et al.}, this would lead one to
conclude that the state preparation fidelity is $ F =
\frac{1}{2}(P_\up + P_\down +A)= 0.75$, even though its actual
value is only $0.5$. This shows clearly how the contribution by
unwanted matrix elements may corrupt the data  for  two-particle
entangled states.

We conclude that this experiment does not provide  evidence of
three-particle entanglement. In order to exclude the contribution
by undesired matrix density elements in the experimental
determination of $\mbox{Re~} \rho_{\uparrow \downarrow}$, another
experimental procedure is needed, e.g.\  an analog of the methods
discussed above, or a test of Conditions \emph{A} and/or \emph{B}
is needed to warrant such a claim.

  \section{Conclusion}

Experimental evidence for $N$-particle entanglement for
$N$-particle states requires stronger conditions than a mere
violation of
 local
realism. $M$-particle entangled states, with $M<N$,  have to be
excluded as well. This leaves a loophole in recent experimental
claims of evidence for multiparticle entangled states.  We have
reviewed two experimentally testable conditions which are
sufficient to close this loophole, and analyzed three recent
experiments to see whether they meet these conditions.
Unfortunately, this is not the case. Hence,\forget{on the basis of
these conditions and because of the indicated loophole problem,}
we conclude that the question remains unresolved whether these
experiments provide confirmation of three-particle entanglement.
However, we have proposed modifications of the experimental
procedure that would make such confirmation possible. We hope that
further experimental tests of $N$-particle entanglement (e.g. the
recently published
 \cite{PAN2}), will take account of the specific requirements
needed to test conditions such as \emph{A} and \textit{B}
discussed above.

\subsection*{Acknowledgments}
The subject of this work was originally suggested  to one of the
authors (M.S.) by Sandu Popescu.
 We are very grateful to him for helpful comments,
and also to Harvey Brown and Dik Bouwmeester for stimulating
discussions and hospitality.

\subsection*{Appendix A}

The data obtained in the experiment of Bouwmeester \emph{et al.}
may be summarized as follows: (i):\ The measurements in the $zzz$
basis give a value of $12:1$ for the ratio between the desired
outcomes and the remainder. This means that \be
\bra{\up\up\down}\rho\ket{\up\up\down} =
\bra{\down\up\up}\rho\ket{\down\up\up} = 0.4  \label{a1}\ee
 and
\be\bra{\up\up\up}\rho\ket{\up\up\up} = \cdots =
\bra{\down\down\down}\rho \ket{\down\down\down} =0.033
\label{a2}\ee
 for the
remaining six outcomes.

(ii) The measurements performed in the $xxx$ directions
determined the probability  of $\op{P}^{(1)}_+ \otimes
\op{P}^{(2)} \otimes \op{P}^{(3)}_\pm $. The experimental results
are depicted in Fig.~2 of Ref.~\cite{BOU1998}, and show a
difference between the $\pm$ settings which is about 75\% of the
expected difference in the desired state $\ket{\psi_{\rm B}}$.
 Hence,
 \BE
 \Tr \rho \, \op{P}^{(1)}_+ \otimes \op{P}_-^{(2)} \otimes \sigma^{(3)}_x &
 =&
 \Tr \rho\,  \op{P}^{(1)}_+ \otimes \op{P}_-^{(2)} \otimes
\left(\op{P}^{(3)}_+ - \op{P}^{(3)}_-  \right)   \nn\\
 &=& \frac{3}{4}\bra{\psi_{\rm
B}}\op{P}^{(1)}_+ \otimes \op{P}_-^{(2)} \otimes \sigma^{(3)}_x
\ket{\psi_{\rm B}} \nn\\
& = &
%-  \frac{3}{4}\frac{1}{4} =
-\frac{3}{16}. \label{a3}\EE

(iii) In a  control measurement, the setting of the polarizer for
the first particle was rotated to the $+z$ direction. This
measurement thus determines the value of $\op{P}^{(1)}_{\up}
\otimes \op{P}^{(2)} \otimes \op{P}^{(3)}_\pm$. In this case, no
interference (i.e., no difference between  the $\pm$ setting for
particle three) was observed. This gives the constraint
 \be \Tr \rho \op{P}^{(1)}_{\up} \otimes
\op{P}^{(2)} \otimes \sigma^{(3)}_x  = 0 . \label{a4}\ee

We now show how most of these results may be reproduced by a
simple two-particle entangled state. Consider the state
 \be \label{W}
  W =
\alpha  \op{P}^{(2)}_- \otimes \op{P}_{S}^{(13)}  +
\frac{1-\alpha}{2} \left( \op{P}_{\ket{\up\up\down}} +
\op{P}_{\ket{\down\down\up}} \right),\ee where $\op{P}_{S}^{(13)}$
is the projector on the singlet state $\frac{1}{\sqrt{2}} (
\ket{\up\down} - \ket{\down\up})= \frac{1}{\sqrt{2}} ( \ket{+-} -
\ket{-+})$.

Using this state (\ref{W}), one finds \be \Tr W
\op{P}^{(1)}_{\up} \otimes \op{P}_-^{(2)} \otimes \sigma^{(3)}_x
= 0  \label{a5}\ee in agreement with Eq.~(\ref{a4}). Moreover, \be
\Tr W \op{P}^{(1)}_+ \otimes \op{P}^{(2)} \otimes \sigma^{(3)}_x
= - \frac{\alpha}{2},\ee which gives agreement with Eq.~(\ref{a3})
for $\alpha = 3/8$. Finally, using this choice for $\alpha$ we
find \be\bra{\up\up\down}W\ket{\up\up\down} =
\bra{\down\up\up}W\ket{\down\up\up} = \frac{13}{32} \approx 0.41,
\ee which is sufficiently close to Eq.~(\ref{a1}).

 The only aspect in which the state (\ref{W}) fails to reproduce the
 experimental data is in the constraint
(\ref{a2}). Instead, the state $W$ gives \be
\bra{\up\down\down}W\ket{\up\down\down} =
\bra{\down\up\up}W\ket{\down\up\up} = \frac{3}{32} \approx
 0.09, \ee
 \be
\bra{\up\up\up}W\ket{\up\up\up}  = \bra{\up\down\up}W
\ket{\up\down\up} =
 \bra{\down\up\up}W
\ket{\down\up\up} = \bra{\down\down\down}W\ket{\down\down\down}
=0. \ee

Of course, the fit of the experimental data might be improved by
varying some parameters of the state (\ref{W}) or utilizing the
margins offered by the finite measurement accuracies. However,
the purpose of this calculation is not to claim that  all these
data may consistently be reproduced by two-particle entangled
state. Rather, we wish to point out that one may approximate  the
data unexpectedly closely, so that a serious quantitative test is
needed before one may claim that these data confirm three-particle
entanglement.

\subsection*{Appendix B}
The two ``Bell signals" measured in the experiment of
Rauschenbeutel \emph{et al.} correspond to
    $ \langle\op{B}_+(\phi)\rangle = \Tr \rho \, \sigma^{(1)}_{x} \otimes  \op{P}^{(2)}_{+} \otimes
    \sigma^{(3)}_{\phi}$
and $\langle \op{B}_-(\phi)\rangle  =\Tr \rho\, {\sigma^{(1)}_{x}}
\otimes \op{P}^{(2)}_{-} \otimes
    \sigma^{(3)}_{\phi}$
where $\op{P}^{(2)}_{\pm }$ are projectors on the ``up'' and
``down'' states for spin in the $x$ direction for particle 2. It
is, however, more convenient to deal with their difference, i.e.,\
$\langle \op{ B}_+(\phi)\rangle - \langle\op{B}_-(\phi) \rangle =
\Tr \rho\, \sigma^{(1)}_{x} \otimes \sigma^{(2)}_x \otimes
    \sigma^{(3)}_{\phi }$.
Let us label the eight basis vectors $\ket{\up\up\up}$,
$\ket{\up\up\down}$, $\ket{\up\down\up}$, $\ket{\up\down\down}$,
$\ket{\down\up\up}$, $\ket{\down\up\down}$, $\ket{\down\down\up}$,
$\ket{\down\down\down}$, consecutively by $1,\dots, 8$.
     A straightforward calculation yields
$ \langle \op{B}_+(\phi) - \op{B}_-(\phi)\rangle = 2|\rho_{72}|
\cos(\phi+\varphi_{72})+2|\rho_{54}|\cos(\phi+\varphi_{54})+2|\rho_{36}|\cos(\phi+\varphi_{36})+2|\rho_{18}|\cos(\phi+\varphi_{18})$
where $\rho_{72}= \rho_{27}^{\star}=|\rho_{72}|
exp(i\varphi_{72})$ and similarly for the other matrix elements.

In a worst-case analysis, all the phase factors such as
$\varphi_{72}$ are chosen equal to $0$ and $|\rho_{54}|$,
$|\rho_{36}|$ and
   $|\rho_{18}|$ should be given  their maximal values compatible
with  the measured populations given in Eq.~(\ref{tab}). These
maximal values are obtained from   the following worst-case
decomposition of the unknown density matrix: $\rho= \alpha \sigma
+\beta\tau  + \gamma\upsilon  +\delta\omega $ with $\sigma,
\tau$, and $\upsilon$ the density matrices of the entangled
states $ 1/\sqrt{2}(\ket{\uparrow\uparrow\uparrow} +
       \ket{\downarrow\downarrow\downarrow})$, $ 1/\sqrt{2}(\ket{\uparrow\downarrow\downarrow} +
       \ket{\downarrow\uparrow\uparrow})$ and $ 1/\sqrt{2}(\ket{\downarrow\uparrow\downarrow} +
       \ket{\uparrow\downarrow\uparrow})$ respectively.
$\omega$ is  an arbitrary density matrix, whose off-diagonal
matrix elements, however, are assumed to have zero entries where
any of the three other states $\sigma, \tau$, and $\upsilon$
 has nonzero entries. Using this decomposition, it follows
that $ |\rho_{18}| =\alpha/2$, $|\rho_{54}|= \beta/2$ and
$|\rho_{36}|=\gamma/2$.

However, since $\sigma_{18} = \sigma_{11} =\sigma_{88}$, and
similar relations for $\tau$ and $\upsilon$,    the fractions
$\alpha, \beta$, and $\gamma $ also contribute to
 the populations  $\rho_{ii}$  of the total state,
 whose measured values are collected above in table (\ref{tab}).
The maximal values compatible with these measured populations
$\rho_{ii}$  are:
  $ \alpha/2 = 0.03 \pm 0.01,
   \beta/2 = 0.04 \pm 0.01, \gamma/2 = 0.06\pm 0.01$
   and the maximal value of $w$ is
   thus $ w= \alpha + \beta
   +\gamma = 0.26 \pm 0.04$,
 and  $2\rho_{72}= A-w = 0.28 \pm0.04 - 0.26 \pm0.03 =0.02\pm0.05$.


\begin{thebibliography}{99}

\bibitem{BOU1998}D.~Bouwmeester, J.-W.~Pan, M.~Daniell, H.~Weinfurter, and
A.~Zeilinger, Phys.\ Rev.\ Lett.\ {\bf 82}, 1345 (1999).

\bibitem{RAUSCH} A.~Rauschenbeutel, G.~Nogues, S.~Osnaghi, P.~Bertet, M.~Brune, J.~Raimond, and
S.~Haroche, Science {\bf 288}, 2024 (2000).

\bibitem{SACKETT} C. A.~Sackett, D.~Kielpinski, B. E.~King, C.~Langer, V.~Meyer,
C.J.~Myatt, M.~Rowe, Q. A.~Turchette, W. M.~Itano, D. J.~Wineland,
and C.~Monroe, Nature {\bf 404}, 256 (2000).

\bibitem{PAN} J.-W. Pan, D.~Bouwmeester, M.~Daniell, H.~Weinfurter, and
A.~Zeilinger,  Nature {\bf 403}, 515 (2000).

\bibitem{PAN2} J.-W.~Pan, M.~Daniell, S.~Gasparoni, G.~Weihs, and
A.~Zeilinger, Phys.\ Rev.\ Lett.\ {\bf 86}, 4435 (2001).

\bibitem{POPESCU} S.~Popescu,  Phys.\ Rev.\ Lett.\ {\bf 74}, 2619
(1995)

\bibitem{LEWENSTEIN}M.~Lewenstein, B.~Kraus, P.~Horedecki, and J.I.~Cirac,
Phys.\ Rev.\ A {\bf 63} 044304 (2001).

\bibitem{GISIN2} N.~Gisin, Phys.\ Lett.\ A {\bf 154}, 201 (1991).

\bibitem{POPROHR} S.~Popescu, D.~Rohrlich, Phys.\ Lett.\ A
{\bf 166}, 293 (1992).
\bibitem{MERMIN} N. D.~Mermin, Phys.\ Rev.\ Lett.\ {\bf 65}, 1838 (1990).

\bibitem{ARDEHALI}M.~Ardehali, Phys.\ Rev.\ A {\bf 46}, 5375 (1992).

\bibitem{GISIN} N.~Gisin and H.~Bechmann-Pasquinucci, Phys.\ Lett.\ A {\bf 246}, 1  (1998).

\bibitem{CHSH} J. F.~Clauser, M. A.~Horne, A.~Shimony, and R. A.~Holt,
Phys.\ Rev.\ Lett.\ {\bf 23}, 880 (1969).

\bibitem{CIRELSON}
B. S.~Cirel'son, Lett.\ Math.\ Phys.\ {\bf 4}, 93 (1980).

\forget{\bibitem{SVETLICHNY}  G.~Svetlichny, Phys.\ Rev.\ D {\bf
35}, 3066 (1987).


\bibitem{Sprivate} G.~Svetlichny, private communication.
}

\bibitem{BOU2000}

D.~Bouwmeester, J.-W.~Pan, M.~Daniell, H.~Weinfurter, and
A.~Zeilinger, ``Multi-Particle Entanglement'' in \emph{the Physics
of Quantum Information}, edited by  D.~Bouwmeester, A.~Ekert, and
A.~Zeilinger (Springer, Berlin,
 2000) pp.~197--209.


\bibitem{GHZ} D. M.~Greenberger, M. A.  ~Horne, A.~Shimony and A.~Zeilinger,
Am.\ J.\ Phys.\ {\bf 58}, 1131 (1990). 

\forget{\bibitem{POPESCU2} S.~Popescu, private communication.}


\end{thebibliography}
\end{document}